\documentclass[aps,twocolumn]{revtex4-1}
\usepackage{graphicx}
\usepackage{amsmath}
\usepackage{dcolumn}
\usepackage{bm}
\input epsf

\begin{document}

\title{Interacting dark matter and cosmic acceleration}
 
\author{V\'ictor H. C\'ardenas$^1$}
\email{victor.cardenas[at]uv.cl}


\author{Samuel Lepe$^2$}
\email{samuel.lepe[at]pucv.cl}

\affiliation{$^1$Instituto de F\'{\i}sica y Astronom\'ia, Universidad de
Valpara\'iso, Gran Breta\~na 1111, Valpara\'iso, Chile}


\affiliation{$^2$Instituto de F\'{\i}sica, Pontificia Universidad Cat\'{o}lica de
Valpara\'{\i}so, Avenida Brasil 2950, Valpara\'{\i}so, Chile}

\begin{abstract}
We study the effect of an explicit interaction between two scalar fields components describing dark matter in the context of a recent proposal framework for interaction. We find that, even assuming a very small coupling, it is sufficient to explain the observational effects of a cosmological constant, and also overcome the problems of the $\Lambda$CDM model without assuming an exotic dark energy.
\end{abstract}


\maketitle

\section{Introduction} 

In the context of the standard model of cosmology, the simplest way we can describe the observations that type Ia supernova are dimmer than expected \cite{sn1a} is by introducing -- by hand -- a cosmological constant, leading to the claimed accelerated expansion and to establish the so far successful Lambda Cold Dark Matter (LCDM) model.

Although this model agreed with almost every observational test, from a theoretical point of view the model can not be taken seriously. First of all, assuming that this model is valid requires us to accept that we live right in a very special time in the history of the universe, something like (again) positioning in the center of the universe (this time including the temporal coordinate). It can be fortuitous, but then this should be consider a huge ``cosmic coincidence''. Another worried around the LCDM model is lambda itself, $\Lambda$. What is it? Why it has today this particular value of $ 1.19 \times 10^{-52} $m$^{-2}$ ? Does $\Lambda$ evolve with time? All these questions (and some more) drives the quest for new ideas that replace $\Lambda$ with something else, to describe what we observe but without the theoretical problems (or worries) we mentioned before. We have named this ``something else'' component, dark energy (DE) \cite{dereview}. Proposals that try to shed light into this problem are those assuming the existence of a quintessence field component \cite{quinta}, somehow a dynamical cosmological constant, and also models where the gravitational theory is modified \cite{modgrav}.

Among the tricks that have been proposed to alleviate the cosmic coincidence problem is to assume that DE (whatever it is) is coupled to dark matter (DM). This is appealing because both dark components are detected only by their gravitational effects, and so they can be confused and is not easy to discriminate each component \cite{kunz}. The usual way to express an interaction between DE and DM \cite{review}, \cite{murgia:2016},\cite{Salvatelli:2014zta},\cite{Pourtsidou:2013nha},\cite{Velten:2014xca},\cite{Abdalla:2014cla},\cite{Erdem:2016hqw} is introducing a new term $Q$ such that
\begin{align}
  \dot{\rho}_{m} + 3H (\rho_{m} + p_{m}) &= -Q \,, \label{eq:matter_dm} \\
  \dot{\rho}_{de} + 3H (\rho_{de}+ p_{de}) &= Q \,, \label{eq:matter_de}
\end{align}
where a dot denotes differentiation with respect to time, $H$ is the Hubble rate, $\rho_{m}$ and $\rho_{de}$ are the energy densities, $p_{m}$ and $p_{de}$ are the pressures. If $Q>0$ DM releases energy into DE, while for $Q<0$ the energy flows in the opposite direction. In the literature it is usual to take the function $Q$ be proportional to the energy densities, as $Q=3H\gamma \rho_{m}$ for example. Fixing $Q$ enable us to completely solve the system and find the solutions $\rho_{m}(a)$ and $\rho_{de}(a)$.

To the DE problem, we have to add the DM problem. Although a puzzle, the DM problem is of a different nature compared to the DE one. We have particle candidates that are under search and there is a certain consensus that this can be solved in the next years. This component, although of a non-baryonic nature, is perfectly possible to exist, beyond the standard model of particle physics. There are no weird features (such as negative pressure, for example) that need to be invoked to hold a model.

A huge effort has been made looking for evidence of non-baryonic DM through basically three ways: indirect detection -- when DM particles produce Standard Model particles (photons, electron/positrons, neutrinos etc) -- by direct detection methods -- when Standard Model particles recoil from collisions with invisible DM -- and from colliders -- where collisions of Standard Model particles may produce missing energy or decay products   \cite{Klasen:2015uma}. Direct-detection experiments rely on the scattering of dark-matter particles
from the halo of the Milky Way in a detector on Earth.
The direct-detection rate depends on the local dark-matter density, currently
estimated to be $\rho = 0.39 \pm 0.03$ GeV cm$^{−3}$.
Dark matter cannot only be detected directly in dedicated experiments
searching for nuclear recoils from the scattering of dark-matter particles or
produced in particle accelerators such as the LHC, but it can also reveal
its existence indirectly. The total number of dark-matter particles does not
change significantly after freeze-out in the early universe, but their spatial
distribution changes considerably during structure formation. The very selfannihilation
that plays a central role in this freeze-out can give rise to a
significant flux of γ-rays, neutrinos, and even antimatter such as antiprotons
and positrons, especially in regions with large dark-matter density.

For example, recently there is interest in exploring the astrophysical consequences of an explicit interaction between DM (whatever it is) with radiation. In \cite{Schewtschenko:2014fca,Boehm:2014vja,Schewtschenko:2015rno} the authors find that even a small interaction is sufficient to alleviate the small satellite problem, for example.

In this letter we want to explore the consequences of a small interaction between two DM components (none of them being exotic) at cosmological scale using a slight modification to the interaction framework we have described above.

\section{Coupling dark matter and analysis} 

If DM is a weakly interacting massive particle (WIMP), the interactions are essential to obtain the relic abundance that support the whole scenario, the so called ``WIMP miracle''. Also the interactions between DM particles with those of the standard model (SM) are essential to the extended campaigns of detection (direct and indirect) of DM using observatories in the ground and in space. Further, as it was mentioned in the previous section, there are increasing evidence for the astrophysical consequences of the interaction between DM and photons, not only as can be seen in the power spectrum of the cosmic microwave background radiation \cite{Wilkinson:2013kia}, \cite{Dolgov:2013una}, but also in the context of the small scale problem such as the small satellite problem \cite{Boehm:2014vja}, \cite{Schewtschenko:2014fca}, \cite{Schewtschenko:2015rno}. Then it is natural to study their consequences at cosmological level.

Let us study the background cosmic evolution assuming an explicit interaction between two species of DM particles, one indicated by an $m$ subscript and other by $x$. No cosmological constant term or an exotic component is introduced. Gravity is described by general relativity (GR). On general grounds the system can be described by a Lagrangian of the type
\begin{equation}
 \mathcal{L}_{\rm tot} = \mathcal{L}_{\rm GR} + \mathcal{L}_{\rm m} + \mathcal{L}_{\rm x} + \mathcal{L}_{\rm int} \,,
  \label{eq:Lag_tot}
\end{equation}
where $\mathcal{L}_{\rm GR}$ is the Einstein-Hilbert Lagrangian and the DM Lagrangians $\mathcal{L}_{\rm m}$ and $\mathcal{L}_{\rm x}$ are assumed to be of the type used in the LCDM model, i.e. the Lagrangian {\it of the free} DM {\it with no interaction}. The interaction between both DM -- the $m$ and the $x$ components -- is introduced as is well known in field theory, through an explicit Lagrangian term $\mathcal{L}_{\rm int}$. From (\ref{eq:Lag_tot}) the field equations are
\begin{gather}
  3H^2 = \rho_{m} + \rho_{x} + \rho_{\rm int} \,, \label{eq:Friedmann} \\
  2\dot{H} + 3H^2 = -{p}_{m} - {p}_{x} - {p}_{\rm int} \,, \label{eq:acceleration} \\
    \dot{\rho}_{m} + 3H (\rho_{m} + {p}_{m}) = {Q}_{m} \,, \label{eq:cons_dm} \\
    \dot{\rho}_{x} + 3H (\rho_{x} + {p}_{x}) = {Q}_{x} \,, \label{eq:cons_r}
\end{gather}
Moreover the total energy conservation implies
\begin{equation}
  \dot{\rho}_{\rm int} + 3H (\rho_{\rm int} +{p}_{\rm int}) + {Q}_{m} + {Q}_{x} = 0 \,,
    \label{eq:energy_constr}
\end{equation}
showing that the new functions are actually constrained. The system (\ref{eq:Friedmann}-\ref{eq:cons_r}) was written before in \cite{Tamanini:2015iia}. In the system (\ref{eq:Friedmann} -\ref{eq:cons_r}) the $Q_i$ functions are obtained from the variation of $\mathcal{L}_{\rm int}$ respect to the i-th degrees of freedom, so they are in general different functions. In the next section we study some examples solutions.

The analysis of the theoretical predictions are also complemented with a statistical study. In this work we concentrate in using data from Type Ia supernova to constraint the values of the parameters in the models. In particular we use the type Ia supernova data from the Pantheon set \cite{pantheon}. This sample consist in 1048 spectroscopically confirmed SNIa in the range $0.01 < z <2.3$. We compute the residuals $\mu - \mu_{th}$ an minimize the quantity
\begin{equation}\label{chi2jla}
\chi^2 = (\mu - \mu_{th})^{T} C^{-1}(\mu - \mu_{th}),
\end{equation}
where $\mu_{th} = 5 \log_{10} \left( d_L(z)/10pc\right) $ gives the theoretical distance modulus, $d_L(z)$ is the luminosity distance, $ C $ is the covariance matrix released in
\cite{pantheon}, and the observational distance modulus takes the form
\begin{equation}\label{mujla}
\mu = m - M + \alpha_1 X - \alpha_2 Y,
\end{equation}
where $ m $ is the maximum apparent magnitude in band B, $ X $ is related to the widening of the light curves, and $ Y $ corrects the color. usually, the cosmology -- specified here by $\mu_{th}$ -- is constrained along with the parameters $ M $, $\alpha_1  $ and $\alpha_2$. The analisys is performed using a public code known as \texttt{emcee} \cite{emcee}. This is a stable, well tested Python implementation of the affine-invariant ensemble sampler for Markov chain Monte Carlo (MCMC) proposed by Goodman \& Weare \cite{GW2010}. 


\section{Some model examples} 

\subsection{Symmetric model}

Let us study first a simple model for the interaction between two DM components that we shall denotate with subscripts $m$ and $x$. Let say $Q_{m} = -3H\alpha \rho_{m}$ and $Q_{x} = -3H\beta \rho_{x}$, expressions already used in the literature. Notice that these two factors do not need to be small, neither equals. We assume that $\alpha$ and $\beta$ are different from zero if some interaction operates between these components.

From Eqs. (\ref{eq:cons_dm}) and (\ref{eq:cons_r}) we obtain
\begin{equation}\label{sol1}
\rho_{m}=\rho^0_{m} a^{-3(1+\alpha)},\hspace{0.5cm} \rho_{x}=\rho^0_{x} a^{-3(1+\beta)},
\end{equation}
where we have assumed that both species have {\it free} Lagrangian of the dust type, i.e., with $p_m=0$ and $p_x=0$ as equation of state. From a variational point of view, as the one advocated in the previous section, the stress energy density of the fluid is obtained from
\begin{equation}
T_{\mu \nu} = - 2 \frac{\delta (\sqrt{-g} \mathcal{L})}{\sqrt{-g} \delta g^{\mu \nu}}=- \frac{2 \delta \mathcal{L}}{ \delta g^{\mu \nu}} +  g_{\mu \nu} \mathcal{L} .
\end{equation}
If we assume that $\mathcal{L}_{int} $ {\it does not contains kinetic terms, } then $ \delta \mathcal{L}_{int}/\delta g^{\mu \nu} =0$, and the stress energy tensor associated with the interaction can be written as $T^{\mu} _{\nu} = \delta ^{\mu} _{\nu} \mathcal{L}_{int}$,
that interpreted in the context of a perfect fluid with energy density $\rho_{int} = \mathcal{L}_{int}$, give rises naturally to a component with EoS $p_{int} = -\rho_{int}$. Using together this relation and Eq.(\ref{sol1}) into Eq.(\ref{eq:energy_constr}) we get
\begin{equation}\label{sol2}
\rho_{int} = \rho^0_{int} + \frac{\alpha \rho^0_m}{1+\alpha} (1-a^{-3(1+\alpha)}) +  \frac{\beta \rho^0_x}{1+\beta} (1-a^{-3(1+\beta)}),
\end{equation}
where $\rho^0_{int} = \rho_{int}(a=1) $. From here is evident that as interactions turn off in this model, this implies that $\rho^0_{int}$ should go to zero. In fact, from (\ref{sol2}) replacing $\alpha = \beta =0 $ leads to $\rho^0_{int} = 0$. Of course, this happens only when the parameters are exactly zero $\alpha = \beta =0$.

Notice that replacing (\ref{sol2}) in (\ref{eq:Friedmann}) we get for $E(z)=H/H_0$

\begin{eqnarray}\nonumber
E^2(z) = 1 + \frac{\Omega_m}{1+\alpha} \left((1+z)^{3(1+\alpha)} - 1 \right)+ \\ \label{modFried}
+ \frac{\Omega_x}{1+\beta} \left( (1+z)^{3(1+\beta)} - 1 \right), 
\end{eqnarray}
where we have defined $\Omega_i = \rho^0 _i/3H^2_0$. At this point we would like to point out the following. Suppose that the parameters $\alpha, \beta \ll 1 $ are small (but not zero), then by expanding (\ref{modFried}) in series, we find that at zero order the Hubble function approaches:
\begin{eqnarray}\label{modFried2}
E^2(z) \simeq  1 - \Omega_m - \Omega_x + (\Omega_m + \Omega_x)(1+z)^{3},
\end{eqnarray}
an expression that corresponds exactly to the LCDM flat model. This means that it is enough to have two dust-like constituents in interaction, no matter how small they are (but non-zero), the resulting cosmological model is very similar to a model without interaction between dust and a cosmological constant.

Although we know this model is a very simple one, we want to test it against observational data, to see to what extend the data adjust the values of the parameters (here $\alpha$ and $\beta$) away from the LCDM model. 

\begin{figure}[h!]
\centering
\includegraphics[width=8cm]{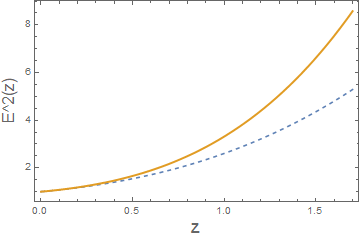}
\caption{Here we reconstruct both (\ref{modFried}) (continuous line) and (\ref{modFried2}) (dahed line) using the best fit values. As is evident, for $z<0.5$ both functions are almost identical, but they start to differs from $z>0.5$.} \label{fig: simetric}
\end{figure}
Because both contributions have the same form (see (\ref{modFried})), there is no point in use two set of different parameters to confront the data. Then we assume that $\Omega_m = \Omega_x = \Omega$ will be the density parameter of both contributions, and that $\alpha = \beta = \gamma$ will be the interaction constant for both contributions too. We know that this choice clearly select a very special type of solution of the system, but we just want to explore how the model behaves once we constraint against the data. 
\begin{figure}[h!]
\centering
\includegraphics[width=8cm]{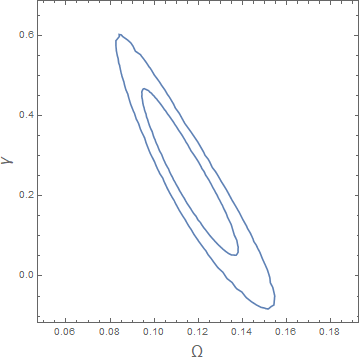}
\caption{Here we display the confidence contours -- at one and two sigmas -- for the two free parameters in the fit: $\Omega$ and $\gamma$ defined in the text. Notice that the data suggest a non zero $\gamma$ at $1 \sigma$ but at $2\sigma$ does not exclude the zero value.  } \label{fig: contour}
\end{figure}
The best fit values obtained in this way are $\Omega = 0.11 \pm 0.03$ and $\gamma = 0.25 \pm 0.15$. Because the ``effective'' density parameter is the sum $\Omega_m + \Omega_x = 2\Omega$ then our model predicts a density parameter for dark matter of $2\Omega = 0.22 \pm 0.06$. In Fig.(\ref{fig: simetric}) we display both (\ref{modFried}) and (\ref{modFried2}) using the best fit values. For small redshift, $z<0.5$ both curves are almost identical, but they differs appreciably for larger ones. 

Let us discuss some points on $\rho_{int}$. Based on the solution (\ref{sol2}) and assuming that $1+\gamma >0$ we get
\begin{equation}\label{eq14}
\rho_{int}\left( z\rightarrow -1\right) \rightarrow \rho_{int}^0 +\frac{\gamma} {1+\gamma}\left( \rho_{m}^0 + \rho_{x}^0 \right),
\end{equation}
so if we want to keep $\rho_{int} \geq 0$ at any moment in the future, we have to ensure that the right hand side of (\ref{eq14}) must be $\geq 0$. For small values of $\alpha$ and $\beta$ (or $\gamma$ in this special case) this means that $\rho_{int}^0 + \alpha \rho_{m}^0 + \beta \rho_x^0 \geq 0$.

Into the past, we would like to have  $\mathbf{\rho }_{int}\left( z\rightarrow
\infty \right) > 0$, so from (\ref{sol2}) we find that 
\begin{eqnarray}
\rho_{int} \rightarrow \rho^0_{int} - \frac{\alpha z^{3(1+\alpha)}\rho^0_m}{1+\alpha}
- \frac{\beta z^{3(1+\beta)} \rho^0_x}{1+\beta},
\end{eqnarray}
a result that seems difficult to achieve unless one of the parameters would be small but negative.

\subsection{Asymmetric model}

Let us assume the asymmetric interaction functions as
\begin{equation}\label{asymmetry}
    Q_m=-3H\alpha \rho_m, \hspace{1cm} Q_x=-3H\beta(\rho_m +\rho_x)
\end{equation}
Replacing the first one in Eq.(\ref{eq:cons_dm}) we obtain
\begin{equation}\label{rom2}
\rho_{m}(a)=\rho^0_{m} a^{-3(1+\alpha)}.
\end{equation}
From (\ref{eq:cons_r}) and the second one in (\ref{asymmetry}) we get


\begin{equation}\label{rox2}
    \rho_x(a)=\rho_x^{0}a^{-3(1+\beta)}+ \beta \rho_m^{0} \frac{\left( a^{-3(1+\beta)}-a^{-3(1+\alpha)}\right)}{\beta-\alpha}
\end{equation}
Following the previous case, we now write and solve the equation for $\rho_{int}$. From (\ref{eq:energy_constr})
\begin{equation}
\frac{d\rho_{int}}{dt}=3H\alpha \rho_m + 3H\beta (\rho_m + \rho_{x}).
\end{equation}
Using the expressions (\ref{rom2}) and (\ref{rox2}) we find 

\begin{equation}\label{rhoint2}
    \rho_{int}=\rho^{0}_{int}+F(a)+G(a),
\end{equation}
where
\begin{equation}\label{fdef}
    F=\left[ -\frac{\alpha^2\rho^0_{m}}{\beta-\alpha} \right]\frac{(1-a^{-3(1+\alpha)})}{1+\alpha},
\end{equation}
and 
\begin{equation}\label{gdef}
G=\left[ \beta\rho^0_{x}+\frac{\beta^2\rho^0_{m}}{\beta-\alpha} \right]\frac{ (1-a^{-3(1+\beta)})}{1+\beta}.
\end{equation}

Let us discuss this solution. Notice that as $z \rightarrow -1$ (or $a \rightarrow \infty$) we get 
\begin{eqnarray}
\rho _{int} \rightarrow  \rho_{int}^0  -\frac{\alpha ^{2}\rho_m^0}{\left(
\beta -\alpha \right) \left( 1+\alpha \right) } + \\
+ \frac{\beta }{%
1+\beta }\left[ \rho_{x}^0 + \frac{\beta \rho_{m}^0}{\beta -\alpha 
}  \right].
\end{eqnarray}
Notice that within this limit and assuming small $\alpha, \beta \ll 1$ we can write
\begin{equation}\label{eq25}
\rho _{int} \rightarrow \rho_{int}^0 +\left( \alpha +\beta \right) \rho _{m}^0 +\rho
_{x}^0 >\rho _{int}^0 ,     
\end{equation}
which certainly implies the condition $( \alpha +\beta) \rho _{m}^0 +\rho
_{x}^0 \geq 0$, that seems easily to achieve.

On the other hand, the limit 
$z\rightarrow \infty $ of (\ref{fdef}) and (\ref{gdef}) tells us that
\begin{eqnarray}
F & \rightarrow & \frac{\alpha ^{2}\rho _{m}^0}{\left(
\beta -\alpha \right) \left( 1+\alpha \right) }
(1+z)^{3\left( 1+\alpha \right) }, \\
\text{ \ \ }G &\rightarrow &-\frac{
\beta }{1+\beta }\left[ \rho _{x}^0 + \frac{\beta \rho _{m}^0}{\beta
-\alpha }  \right] \left( 1+z\right)
^{3\left( 1+\beta \right) },
\end{eqnarray}%
and then
\begin{eqnarray}\nonumber
\rho _{int} \rightarrow \rho_{int}^0 +
\frac{\alpha ^{2}\rho _{m}^0}{\left(
\beta -\alpha \right) \left( 1+\alpha \right) }
(1+z)^{3\left( 1+\alpha \right) }+\\
-\frac{
\beta }{1+\beta }\left[ \rho _{x}^0 + \frac{\beta \rho _{m}^0}{\beta
-\alpha }  \right] \left( 1+z\right)
^{3\left( 1+\beta \right) }.
\end{eqnarray}%
An expression that reduces to the following after considering up to first order terms assuming $\alpha, \beta \ll 1 $
\begin{equation}
\rho _{int} \rightarrow \rho _{int}^0 -\left[ \left( \beta +\alpha \right) \rho _{m}^0
+\beta \rho _{x}^0 \right] \left( 1+z\right) ^{3},
\end{equation}
which may implies that the term in square parenthesis should be $[...] \leq 0$ in order to get a $\rho _{int}\left( z\rightarrow \infty \right) \geq 0$. If we emphasize this point, we can express the condition even more explicitly, $\left( \beta +\alpha \right) \rho _{m}^0
+\beta \rho _{x}^0=0$ or
\begin{equation}\label{eq31}
\frac{\rho _{m}^0}{\rho _{x}^0} = - \frac{\beta}{\alpha + \beta}.
\end{equation}
According to our best fit values for $\alpha$ and $\beta$ both sides are numbers close to one, and within the errors it is satisfied. Notice that both limits points to consistent condition (see (\ref{eq25}) and (\ref{eq31})).

Adding (\ref{rom2}), (\ref{rox2}) and (\ref{rhoint2}) in (\ref{eq:Friedmann}) we find the Hubble function $H(a)$. Evaluating it today we get
\begin{equation}\label{h0def}
    3H_0^2=\rho^0_{m} + \rho^0_{x} + \rho^0_{int},
\end{equation}
as it can be. Using this relation to replace $\rho^0_{int}$, we can write $E=H/H_0$ and find

\begin{eqnarray}\nonumber
E^2=1+\Omega_m\left( -1 + a^{-3(1+\alpha)}+S(a)+R(a)\right) + \\ \label{edzasym}
+ \Omega_x\left( -1 + a^{-3(1+\beta)}+T(a)\right),
\end{eqnarray}
where 
\begin{equation}\label{sdz}
S=\beta \frac{a^{-3(1+\beta)}-a^{-3(1+\alpha)}}{\beta-\alpha},
\end{equation}
\begin{equation}\label{rdz}
R=\frac{\beta^2}{\beta-\alpha} \frac{1 - a^{-3(1+\beta)}}{1+\beta} -\frac{\alpha^2}{\beta - \alpha}\frac{1- a^{-3(1+\alpha)}}{1 +\alpha} ,
\end{equation}
\begin{equation}\label{tdz}
T=\beta \frac{1 - a^{-3(1+\beta)}}{1+\beta},
\end{equation}

In the limit for small $\alpha$, and $\beta$ we find at first order in the parameters that the function $S(z)$ goes like
\begin{equation}\label{sdzlim}
    S(z) \simeq -3 \beta (1+z)^3 \log (1+z) +O(\alpha^2, \alpha \beta, \beta^2)
\end{equation}
the $R(z)$ function evolves as
\begin{equation}\label{rdzlim}
    R(z) \simeq (\alpha + \beta)\left( 1-(1+z)^3\right) + O(\alpha^2, \alpha \beta, \beta^2)
\end{equation}
and finally the $T(z)$ behaves as
\begin{equation}\label{tdzlim}
    T(z) \simeq (\alpha + \beta)\left( 1-(1+z)^3\right)+ O(\alpha^2, \alpha \beta, \beta^2),
\end{equation}
this means that at order zero we get the same result as the previous section: the LCDM limit.

Let us study the behavior of the solution in two limits: the future at $z \rightarrow -1$ and also into the distant past to $z \rightarrow \infty$. Taking the expressions (\ref{sdz}), (\ref{rdz}), and (\ref{tdz}) and looking for the limit $z \rightarrow -1$ we get
\begin{eqnarray}
S &= & \frac{\beta (1+z)^3}{\beta
-\alpha } \left[ \left( 1+z\right) ^{3\beta }-\left( 1+z\right)
^{3\alpha }\right] \rightarrow 0, \\
R &= & \frac{1}{\beta -\alpha } \left( \frac{\beta ^{2}}{1+\beta }-\frac{\alpha ^{2}}{1+\alpha }%
\right) , \\
T &= & \frac{\beta }{1+\beta }
\end{eqnarray}%
and if $\mathbf{\alpha ,\beta <<1}$
which then implies that
\begin{eqnarray}\nonumber
E^2 \rightarrow 1-\Omega _{m} \left[ 1-\left( \frac{1}{\beta -\alpha }\right) \left( \frac{\beta ^{2}}{1+\beta } + \right. \right. \\
- \left. \left. \frac{\alpha ^{2}}{1+\alpha }\right) \right] -\Omega _{x}\left[ 1-\frac{\beta }{1+\beta }\right],
\end{eqnarray}
that after assuming $\alpha, \beta \ll 1$ reduces 
\begin{equation}
 E^{2}\left( z\rightarrow -1\right) \rightarrow 1-\left( \Omega _{m}+\Omega
_{x}\right),   
\end{equation}
as we have anticipated.

Further, as we have obtained above, for small $\alpha, \beta \ll 1$ we get (\ref{sdzlim}), (\ref{rdzlim}) and (\ref{tdzlim}), and then we can write up to first order
\begin{eqnarray}\nonumber
E^{2} = 1 + \Omega_{m}\left[ \left[ \left( \alpha +\beta \right) -1\right] \left( 1-\left( 1+z \right) ^{3}\right) + \right. \\ \nonumber 
\left. -3\beta \left( 1+z\right) ^{3}\log \left( 1+z\right) \right] + \\ 
+\Omega _{x}\left[ \left( \alpha +\beta \right) -1\right] \left( 1-\left( 1+z\right) ^{3}\right),
\end{eqnarray}
in this way we get in the limit $ z\rightarrow -1$ 
\begin{equation}
E^{2}\rightarrow 1-\left( \Omega _{m}+\Omega
_{x}\right) -3\beta \Omega _{m}\left[ \left( 1+z\right) ^{3}\log \left(
1+z\right) \right], 
\end{equation}
where the term in square brackets goes to zero as $z\rightarrow -1$. This means that independent of the value of $\beta$ the $\Lambda$CDM limit (for $z\rightarrow -1$) is restored.

\begin{figure}[h!]
\centering
\includegraphics[width=9cm]{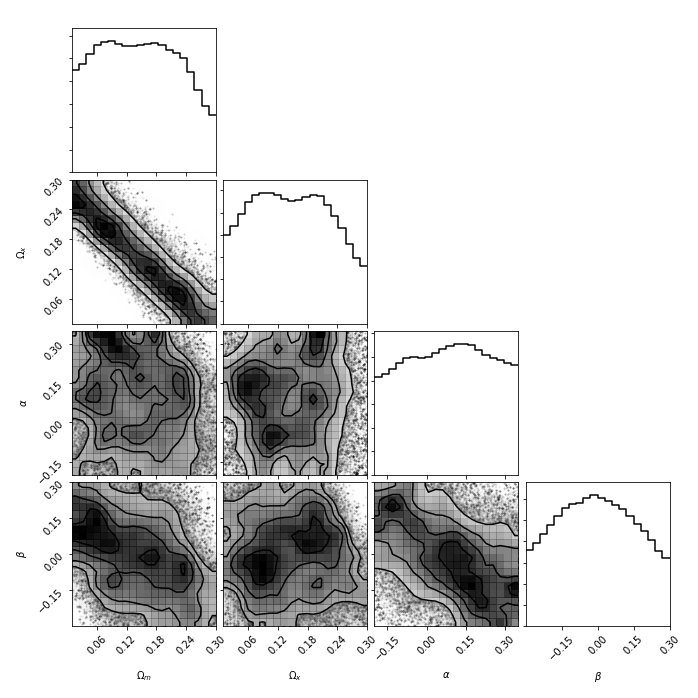}
\caption{Here we show the result of the statistical analysis using the Pantheon sample of type Ia supernovae.} \label{fig: asimetric}
\end{figure}

It is then interesting to see what the analysis using observational data can gives us about this model. Using the latest supernova data \cite{pantheon} as we did in section III, we find the following values as the best fit for the parameters:  $\Omega_m = 0.14 \pm 0.09 $, $\Omega_x = 0.146 \pm 0.085$, $\alpha = 0.09 \pm 0.17$, and $\beta = -0.01 \pm 0.17$. The result of 5000 chains using the \texttt{emcee} code \cite{emcee} is shown in Fig.(\ref{fig: asimetric}).

The data then implies small values for the parameters $\alpha$ and $\beta$ that controls the interaction between these dark matter components. The small values of these values are in agreement with our statement of getting close to the $\Lambda$CDM model in this limit. 

The total dark matter density parameter gives us $ \Omega_m + \Omega_x \simeq 0.28 $ a value that is in agreement with other astrophysical tests. From Fig.(\ref{fig: asimetric}) it is clear the degeneracy between these two dark matter component, but keeping the sum essentially constant. 

\begin{figure}[h!]
\centering
\includegraphics[width=9cm]{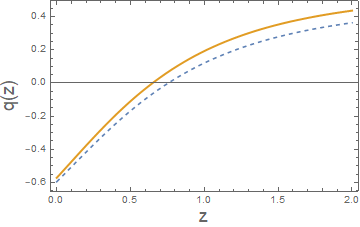}
\caption{Here we display the deceleration parameter for our asymmetric model (dashed line) together with that from the flat LCDM model (continuous line). } \label{fig: qvsz}
\end{figure}

Using these best fit values we can plot the reconstructed deceleration parameter as a function of redshift $z$. We display it in Fig.(\ref{fig: qvsz}) together with the deceleration parameter for the flat LCDM model with $\Omega_m=0.27$. As we can see, both curves essentially follows the same trend with a very small difference in amplitude (of our model compared to that from the LCDM model). The redshift for the transition between deceleration to acceleration is around $z \simeq 0.7-0.8$ for the models.

The best fit values obtained for the free parameters $\alpha$ and $\beta$ are then small enough to obtain a model very close to the LCDM {\it at small redshift}. It is then logic to ask for the performance of this model for large redshifts. To answer appropriately this question it necessarily take us beyond the scope of this work. However, we can try to answer some of the main questions. In this case is imperative to add the contributions for both baryons and radiation that are negligible at small redshift but which are increasingly important as we move to large $z$. However, we can not use the best fit values for $\alpha$ and $\beta$ obtained from the test using SNIa because these are valid for the range where data is available, this is $0.01<z<2.3$. Necessarily the extension of our model beyond redshift 2 is an extrapolation that can not be taken seriously. We need then to test a modified model -- with baryons and radiation -- and test it as we go back in time.

In order to check this, we need an observational probe that constrain the model at large $z$ as the cosmic background radiation (CMB). Following \cite{zaiwuan} we use the three CMB distance priors, the shift parameter $R$, the acoustic scale $l_a$ and the baryon density parameter $\Omega_b h^2$. The details of the computation of the constraints from CMB follows \cite{cardenas} and is explained in the Appendix, however is relevant to explain certain points here. First of all, we have to add baryons and radiation explicitly in the Friedman equation Eq.(\ref{eq:Friedmann}). Because these two components conserved separately, Eq.(\ref{eq:energy_constr}) is not modified.
The best fit values obtained are $\alpha=0.07$, $\beta=-0.05$ with $\Omega_m=0.142$ and $\Omega_x=0.16$. These values were obtained in a joint analysis with SNIa and CMB priors. A detailed study of this model at large redshift model is underway, but we can conclude from these preliminary results that: first, having inserted baryons and radiation, the model is able to fit the data at large redshift with small interaction parameters in agreement with that implied by the SNIa data (at low redshift data). This means that our model is capable to fit simultaneously both small and large redshift data keeping $\alpha, \beta$ small, and with a DM density $\Omega_m +\Omega_x \simeq 0.3$. This conclusion is reinforced from a explicit computation of the age of the universe in this model. By using the Hubble function $H(z)$, with baryons and radiation added, and making use the best fit values for the parameters $\alpha$, $\beta$, $\Omega_m$ and $\Omega_x$, we get
\begin{equation}
    t_0 H_0 = \int_0^{\infty} \frac{dz}{(1+z)E(z)} \simeq 0.9743,
\end{equation}
which gives us an age of the Universe similar to that inferred from the LCDM model.

Another concern would be if the interaction functions here defined growth with redshift spoiling the matter formation era. A simple way to check this, following \cite{referee}, is considering the relative strength of the coupling $f$ for each DM component: $f_m= Q_m/3H\rho_m$ and $f_x=Q_x/3H\rho_x$. In the case of the symmetric model we obtain $f_m=\alpha$ and $f_x=\beta$. This also is obtained in the asymmetric model where $f_m=\alpha$ and $f_x \simeq \alpha$. Because the strength parameters does not grow with redshift, but reach constant small values, we can expect to obtain a similar behavior that those in \cite{Salvatelli:2014zta} where a small interaction parameters not only is possible, but it seems to be needed to obtain a better fit of the process of the structure formation.

\begin{figure}[h!]
    \centering
    \includegraphics[width=8cm]{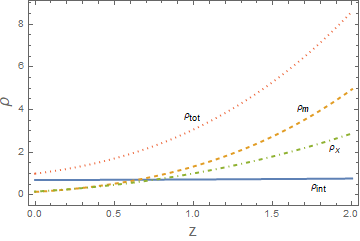}
    \caption{Here we display each energy density component using the best fit values obtained using type Ia supernovae + CMB priors constraints.}
    \label{fig: roes}
\end{figure}

It is also interesting to see the evolution of each energy density component as a function of redshift. This is display in Fig.(\ref{fig: roes}) 
Although $\rho_{int}$ seems to keep a constant value, its value actually is decreasing with redshift until reaching a zero value at $z \simeq 7$ after which it becomes negative. However, we must be cautious with these results, since our physical system is of interaction between two components and therefore the analysis of each one separately, and especially the interaction component, do not make much sense. Physically only the total density makes sense.

\begin{figure}[h!]
    \centering
    \includegraphics[width=8cm]{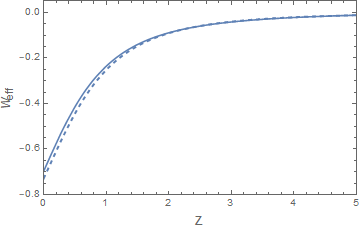}
    \caption{Here we plot the total equation of state for our model (continuos line) in comparison to that of the LCDM model (dashed line).}
    \label{fig: weffs}
\end{figure}

We can also compute the total equation of state of our model using (\ref{eq:Friedmann}) and (\ref{eq:acceleration}). Explicitly we just need to compute
\begin{equation}
\omega _{tot}(z) = \frac{p_{tot}}{\rho_{tot}}=-\frac{\rho_{int}}{\rho_m +\rho_x+\rho_{int}},
\end{equation}
which can be plotted using the best fit values recently obtained. The results is display in Fig.(\ref{fig: weffs}) together with the equivalent to the LCDM model. We notice a very small difference between them.
As we see in the plot, at small redshifts, the effective EoS of our model is slightly larger than that of the LCDM model, however this trends change around redshift $z \simeq 2.1$ after that the $w_{eff}$ of our model remains larger for a while until $z \simeq 4$ where again turns out to be larger. We have explore these changes for large redshift and always their difference (in the total EoS parameter) is minor than $0.03$.

\section{Thermodynamics}

Let us discuss the thermodynamics of the models proposed in previous section. We want to explore to what extend these new models for dark energy are consistent with the laws of thermodynamics.

\subsection{Symmetric case}

The fluid $\rho _{int}$ behaves as $\Lambda $ does in the sense that this contribution satisfy $p_{int}=-\rho _{int}$. The fluid $\rho _{m}$ behaves as one perceiving an effective {\it small} pressure $p_{m}^{eff}=\alpha \rho _{m}$ and also the fluid $\rho _{x}$ as a fluid with $p_{x}^{eff}=\beta \rho _{x}$, where we have defined the effective EoS parameters $\omega _{m}^{eff}=\alpha $ and  $\omega _{x}^{eff}=\beta $. In this way, the coupled system reduce to
\begin{eqnarray}
\dot{\rho}_{m}+3H\left( 1+\alpha \right) \rho _{m} &=&0, \\
\dot{\rho}_{x}+3H\left( 1+\beta \right) \rho _{x} &=&0,
\end{eqnarray}%
that can be studied along the discussion in \cite{CGL} (see also \cite{Maartens:1996vi}). In this context, the temperature of the fluid can be written as
\begin{equation}\label{teff}
T\left( z\right) =T^0 \exp \left( 3\int_{0}^{z}\frac{dz}{1+z}%
\omega ^{eff}\left( z\right) \right) ,     
\end{equation}
then we can write for each component
\begin{equation}
 T_{m}\left( z\right) =T_{m}^0 \left( 1+z\right) ^{3\alpha }%
\text{ \ },\text{ \ }T_{x}\left( z\right) =T_{x}^0 \left(
1+z\right) ^{3\beta },   
\end{equation}
in this way and according to \cite{CGL} if  $\alpha \neq \beta \neq 0$ there is no adiabaticity. This is recovered for $\alpha =\beta =0$.

In general, without assuming an explicit EoS parameter for the fluids, and 
according to the first law
\begin{equation}
TdS=d[(ρ +p)V]-Vdp,    
\end{equation}
we can write for each component
\begin{eqnarray}
T_{m}\frac{dS_{m}}{dt} &=&\frac{d}{dt}[(1+\omega _{m})ρ _{m}V]-\omega
_{m}V\frac{dρ _{m}}{dt}, \\
T_{x}\frac{dS_{x}}{dt} &=&\frac{d}{dt}[(1+\omega _{x})ρ _{x}V]-\omega
_{x}V\frac{dρ _{x}}{dt},
\end{eqnarray}%
which can be rewritten as
\begin{eqnarray}
\frac{T_{m}}{V}\frac{dS_{m}}{dt} &=&\frac{dρ _{m}}{dt}+3H(1+\omega
_{m})ρ _{m}=Q_{m}, \\
\frac{T_{x}}{V}\frac{dS_{x}}{dt} &=&\frac{dρ _{x}}{dt}+3H(1+\omega
_{x})ρ _{x}=Q_{x}
\end{eqnarray}%
where we have set $dV/Vdt=3H$. In this way
\begin{eqnarray}
\frac{T_{m}}{V}\frac{dS_{m}}{dt} &=&Q_{m}\text{ \ \ },\text{ \ \ }\frac{T_{x}%
}{V}\frac{dS_{x}}{dt}=Q_{x} \\
&\Longrightarrow &\frac{T_{m}}{Q_{m}}dS_{m}=\frac{T_{x}}{Q_{x}}dS_{x}.
\end{eqnarray}%
So we can write
\begin{equation}
 d\left( S_{m}+S_{x}\right) =\left( 1+\frac{Q_{m}}{Q_{x}}\frac{T_{x}}{T_{m}}%
\right) dS_{x},    
\end{equation}
and then if
\begin{equation}
d\left( S_{m}+S_{x}\right) =\left( 1+\frac{\alpha }{\beta }\frac{\rho
_{m}T_{x}}{\rho _{x}T_{m}}\right) dS_{x}\neq 0,     
\end{equation}
we know there is no adiabaticity. Replacing what we have found previously for temperatures we get
\begin{equation}
\frac{\rho _{m}T_{x}}{\rho _{x}T_{m}} =\frac{\rho^0_{m}
T_{m}^0 }{\rho^0_{x} T_{x}^0 }\left(
1+z\right) ^{6\left( \alpha -\beta \right) }. 
\end{equation}
In particular we observe that
\begin{equation}
    \frac{\rho _{m}T_{x}}{\rho _{x}T_{m}}\left( z\rightarrow -1\right)
\rightarrow 0
\end{equation}
which implies that
\begin{equation}
dS_{m}\left( z\rightarrow -1\right)
+dS_{x}\left( z\rightarrow -1\right) \rightarrow dS_{x}\left( z\rightarrow
-1\right) ,    
\end{equation}
then its clear that $dS_{m}\left( z\rightarrow -1\right) \rightarrow 0, $ and then also $ dS_{x}\left( z\rightarrow -1\right) \rightarrow 0, $ 
so we restore the adiabaticitcy in this limit (the $\Lambda$CDM limit). Notice that it seems relevant if $\alpha >\beta $ for this conclusion to be correct. However, as we will discuss in brief, there is no real meaning to this inequality due to the symmetry of the model. Now if we assume $\alpha, \beta \ll 1 $ then $ \rho _{m} \left( z \right) \simeq \rho^0_{m} \left( 1+z\right)^{3}$, and $\rho _{x}\left( z \right) \simeq  \rho^0_{x} \left( 1+z\right) ^{3},$
so we find
\begin{equation}
d\left( S_{m}+S_{x}\right) =\left( 1+\frac{\rho^0_{m}
T_{x}^0 }{\rho^0_{x} T_{m}^0 }\frac{
\alpha }{\beta }\left( 1+z\right) ^{6\left( \alpha -\beta \right) }\right)
dS_{x}    
\end{equation}
which implies there is no adiabaticity. So only when $z\rightarrow -1$ it is possible to restore the adiabaticity. However, this is not completely correct, because once we assume that $\alpha >\beta $ or $\alpha < \beta $, we are making a choice about the future evolution of the system. Further, since in this context $\Omega_m$ and $\Omega_x$ are also interchangeable contributions, there is no need to worry about a particular hierarchy choice.

Let us discuss now the case for $\rho_{int}$. From the combination of the first and second law
\begin{equation}
 T_{int}dS_{int} = d[(\rho_{int}+p_{int})V]-Vdp_{int}.   
\end{equation}
Using the EoS for the compnent $p_{int} = -\rho_{int}$ we find
\begin{equation}
T_{int}\frac{dS_{int}}{dt}=V\frac{%
d\rho _{int}}{dt},
\end{equation}
then, using the conservation equation for $\rho_{int}$ we find 
\begin{equation}
 \frac{T_{int}}{V}=-\left( Q_{m}+Q_{x}\right) =3H\left( \alpha
\rho _{m}+\beta \rho _{x}\right) ,   
\end{equation}
from which we conclude
\begin{eqnarray}
\frac{dS_{int}}{dt}>0,
\end{eqnarray}%
and then, according to the relation,
\begin{eqnarray}
\frac{\dot{T}_{int}}{T_{int}} &=&-3H\left( \frac{\partial p_{int}}{\partial
\rho _{int}}\right) =3H,
\end{eqnarray}%
the temperature evolves as $T_{int}\left( z\right) =T_{int}\left( 0\right) \left(
1+z\right) ^{-3},$ so in the future limit $T_{int}\left( z\rightarrow -1\right) \rightarrow \infty$, a result typical of models of DE.

We have to stress here that in the case of $\Lambda$ we have $p=-\rho =-\Lambda $ and obviously $\dot{\rho}=0$ so in this case we can not use the relation $\dot{T}/T=-3H\left( \partial
p/\partial \rho \right) $. As we know \cite{CGL} the temperature associated to $\Lambda $ is zero, while in the present case $p_{int}=-\rho_{int}$ y $\dot{\rho}_{int}\neq 0$, making evident an important difference between these two contributions.

As we mentioned, the previous statement about the temperature $T_{int}\left( z\rightarrow -1\right) \rightarrow \infty $ should not be a surprise for us. In fact, for a generic dark energy component $p_{de}= \omega_{de}\rho_{de}$ with $\omega_{de}<0$ we have  $\dot{T}_{de}/T_{de}=3\left\vert \omega _{de}\right\vert H,$ then the temperature evolves as $T_{de}\left( z\right) =T_{de}\left( 0\right) \left( 1+z\right) ^{-3\left\vert \omega _{de}\right\vert },$ so in the future limit we get $T_{de}\left( z\rightarrow -1\right)
\rightarrow \infty$. In this sense $\rho_{int}\left( z\right) $ plays a better role (a more physically stronger role) as dark energy than those played by $\Lambda $.

It is interesting also to notice that in the context of $\Lambda$CDM, the transition redshift between deceleration/acceleration occurs usually around $z \simeq 0.5$ which is also the redshift from which our exact solution (\ref{modFried}) start to differs from the $\Lambda$CDM limit solution (\ref{modFried2}).

As a summary, we have a very simple model far more physically sound than the cosmological constant, where a very small coupling between dark matter components behaves as $\Lambda$CDM. This is the case for the symmetric model, where the change $\alpha \Longleftrightarrow \beta$ and $\Omega_m \Longleftrightarrow \Omega_x$ left the Hubble function unchanged. In the next sub section we discuss the asymmetric case.

\subsection{Asymmetric case}

Here we discuss the thermodynamics consequences of the asymmetric model previously presented. Let us start rewriting the system of conservation equations for both components. From (\ref{asymmetry}) we have
\begin{eqnarray}
\dot{\rho}_{m}+3H\left( 1+\alpha \right) \rho _{m} &=&0, \\
\dot{\rho}_{x}+3H\left( 1+\beta \left[ 1+\frac{\rho _{m}}{\rho _{x}}\right]
\right) \rho _{x} &=&0.
\end{eqnarray}%
where $\omega
_{m}^{eff}=\alpha $ and $\omega _{x}^{eff}\left( z\right) =\beta %
\left[ 1+\rho _{m}(z)/\rho _{x}(z)\right]$. Explicitly the quotient 
$\rho _{m}/\rho _{x}$
takes the form 
\begin{equation}
\frac{\rho _{m}}{\rho _{x}} =\frac{\rho^0_{m}}{\rho^0_{x}} \frac{\left( 1+z\right) ^{3\left( \alpha -\beta \right) }}{1+
\left( \frac{\beta}{\beta -\alpha} \right) \frac{\rho^0_{m}}{\rho^0 _{x}}  \left[ 1-\left( 1+z\right)
^{3\left( \alpha -\beta \right) }\right] },
\end{equation}
then assuming $\alpha >\beta $ we can take the future limit $z \longrightarrow -1$ we find that $\rho _{m}/\rho _{x}\rightarrow 0$ which implies that $\omega _{x}^{eff}\left( z\rightarrow -1\right) \rightarrow
\beta $. On the other hand, in the limit of the far past $z \longrightarrow \infty$ we get that  $\omega _{x}^{eff}\left( z\rightarrow
\infty \right) \rightarrow \alpha =\omega _{m}^{eff}.$

Now, let us compute explicitly the temperatures. Using the formula (\ref{teff}) this leads to
\begin{eqnarray}
T_{m} &=&T_{m}^0 \left( 1+z\right) ^{3\alpha },
\\
T_{x}  &=&T_{x}^0 \left( 1+z\right) ^{3\beta
}\exp \left( 3\beta \int_{0}^{z}\frac{dz}{1+z}\frac{\rho _{m}\left( z\right) 
}{\rho _{x}\left( z\right) }\right) 
\end{eqnarray}
Following the same steps described in the analysis of the symmetrical case, we have

\begin{eqnarray}\nonumber
d\left( S_{m}+S_{x}\right)  &=&\left[ 1+\frac{Q_{m}}{Q_{x}}\frac{T_{x}}{T_{m}%
}\right] dS_{x}, \\
&=&\left[ 1+\frac{\alpha }{\beta }\frac{\rho _{m}}{\rho _{x}}\left( \frac{1}{%
1+\rho _{m}/\rho _{x}}\right) \frac{T_{x}}{T_{m}}\right] dS_{x}%
,
\end{eqnarray}%
which implies no-adiabaticity. However, as we take the limit $z\rightarrow -1$, then $\rho _{m}/\rho _{x}\rightarrow
0$, then we obtain that both $dS_{m} \rightarrow 0$ and $dS_{x}\rightarrow 0$, and the adiabaticity is restore in this limit.

\section{Towards a field model}

In this section we describe a possible implementation of the model presented in section III. For this, we will use two scalar fields whose free behavior, that is, ignoring interaction between them, behave like dark matter, that is, a dust-like evolution.

It is well known \cite{turner83} that coherent scalar field oscillations with a self interacting potential $ \simeq \phi ^n$, behaves as a contribution whose energy density decay as $a^{-6n/(n+2)}$. For a pure DM contribution, the energy density goes as $a^{-3}$ then the potential would be $V(\phi) = V_0 \phi^{2}$. This is an exact result assuming that no other constituent than the scalar field is present.

Another way to build up a scalar field behaving as DM, is by using the reconstruction scheme. From \cite{reconst} the scalar field potential and kinetic term can be written in terms of the scale factor through the parametric equations
\begin{equation}
U(t) = \frac{3}{8\pi G}\left( H^2 + \frac{\dot{H}}{3}\right), \hspace{0.3cm} \dot{\chi (t)}^2 = -\frac{\dot{H}}{4\pi G} .
\end{equation}
Then by using $a(t) = (t/t_0)^{2/3}$ typically of a dust like contribution, we get $H = 2/(3t)$ and $\dot{H}=-2/(3t^2)$, then from the field equation we get
\begin{equation}
    \dot{\chi} = \frac{1}{\sqrt{6\pi G}t}, \implies \ln t = \sqrt{6\pi G}\chi(t),
\end{equation}
then after we write the scalar field potential
\begin{equation}
    U(t) = \frac{3}{8\pi G}\left( \frac{2}{9t^2}\right) \implies U(\chi) = \frac{e^{\sqrt{6\pi G}\chi}}{12\pi G},
\end{equation}
Then for the case of a dust evolution -- $a(t) \simeq t^{2/3}$ -- the equations leads us to an exponential potential $U(\chi)= U_0 \exp(-\alpha \chi)$, in which $\alpha = \sqrt{6\pi G}$. This is a well know result \cite{Liddle:1998xm}.

In our model then, we consider these two scalar fields $\phi(t)$ and $\chi(t)$ interacting through
\begin{equation}
L=\frac{1}{2}\dot{\phi}^2- V(\phi) + \frac{1}{2}\dot{\chi}^2 - U(\chi) -  \frac{1}{2}g^2 \phi^2 \chi^2.
\end{equation}
The stress energy tensor for the {\it homogeneous free fields} $\phi(t)$ and $\chi(t)$ can be written as those of a perfect fluid with energy density and pressure given by
\begin{equation}\label{dens pres}
\rho_{\phi} = \frac{1}{2}\dot{\phi}^2 + V(\phi), \hspace{0.6cm} p_{\phi} = \frac{1}{2}\dot{\phi}^2- V(\phi)
\end{equation}
The interaction Lagrangian can also be written in a perfect fluid form, but this time the energy density and pressure are
\begin{equation}\label{rho int}
\rho_{\phi \chi}= \frac{1}{2}g^2\phi^2 \chi^2, \hspace{0.6cm} p_{\phi \chi}=- \frac{1}{2}g^2 \phi^2 \chi^2,
\end{equation}
which -- as we have anticipated -- automatically satisfies the cosmological constant equation of state, although the energy density is not constant.

The field equations are: the Friedman equation (\ref{eq:Friedmann}) and (\ref{eq:acceleration}) with pressures and densities defined by (\ref{dens pres}) and (\ref{rho int}) before (where the notation $\rho_{int}=\rho_{\phi \chi}$), and the well known
\begin{equation}\label{phiem}
\ddot{\phi} + 3H\dot{\phi} + V'(\phi) = -g^2 \chi^2 \phi,
\end{equation}
and
\begin{equation}\label{chiem}
\ddot{\chi} + 3H\dot{\chi} + U'(\chi) = -g^2 \phi^2 \chi,
\end{equation}
which are the equivalent to (\ref{eq:cons_dm}) and (\ref{eq:cons_r}), where we can identify 
\begin{equation}\label{qes}
Q_{m} = -g^2 \chi^2 \phi \dot{\phi}, \hspace{1cm} Q_{x} = -g^2 \phi^2 \chi \dot{\chi}.
\end{equation}
From (\ref{qes}) and (\ref{rho int}) is clear that (\ref{eq:energy_constr}) is automatically satisfied.

This model clearly show the way we can built a field model of two DM component in interaction with an evolution similar to the $\Lambda$CDM model. A work in progress in underway where we focus on this specific model.

\section{Discussion}

In this paper we have proposed a family of models for DE consisting in two DM species interacting each other, whose interaction although small, enable us to describe a typical evolution of the $\Lambda$CDM model. The key element is the role accomplished by the energy density associated to the interaction Lagrangian, $\rho_{int}$. Assuming the interaction Lagrangian does not have derivative couplings, the automatic equation of state that this component satisfy is $p_{int}=-\rho
_{int} $ i.e., that of the cosmological constant. However, although this component satisfy this EoS, the energy density evolves (in contrast to $\Lambda$ that keep its value constant), making it a more sound component physically speaking. For example, this does not suffer from the ``coincidence problem'' because the interaction energy density -- which is interpreted here as the equivalent to $\Lambda$  -- emerges from a Lagrangian that connects both DM species from the beginning. This connection also answer our question about the order of magnitude of $\Lambda$. Here the response is in essence because both DM contribution are tied through the {\it interaction} which established the order of magnitude of their contributions. Although obvious, it is also necessary to highlight the fact that we do not need an exotic (negative pressure) component to describe the observations.

In fact, from a thermodynamic point of view, this component behaves more naturally than $\Lambda$, showing a temperature that increases in the future, a behavior typical to other DE models where the EoS parameter varies with redshift, in contrast to the $\Lambda$ behavior where the temperature associated is zero. Furthermore, we have discussed how the non-adiabaticity emerges from the model, clearly due to the manifest interaction, and its future evolution towards adiabaticity. Certainly, a much more physical behavior than the disconnected evolution between $\Lambda$ and the rest of the constituents of the universe that is evident in the $\Lambda$CDM model.

We have also performed a statistical analysis using the latest data set for type Ia supernova (the Pantheon sample \cite{pantheon}) consisting in 1048 data points and its covariance. Although very simplistic -- because we have not added a explicit curvature or a baryonic term or radiation term in the Hubble function -- our models are able to describe successfully the data, with small best fit values for the parameters, being in agreement with the hypothesis of the model. Although the errors are big, the contrast with observational data implies the existence of a large family of models with small $\alpha$ and $\beta$ parameters, that describe an evolution that mimic the $\Lambda$CDM model without the necessity to add an exotic dark component.

\section*{Acknowledgements}
We are grateful to Miguel Angel Cruz for helpful discussions.

\appendix*

\section{}

Here we describe the formulae to use CMB priors to constraint our model. This analysis follows \cite{zaiwuan} and \cite{cardenas}. We use CMB information by using the Planck data \cite{planck} extracted from the analysis performed by \cite{zaiwuan} to probe expansion history up to the last scattering surface. The $\chi^2$ for the CMB data is constructed as
\begin{equation}\label{cmbchi}
 \chi^2_{CMB} = X^TC_{CMB}^{-1}X,
\end{equation}
where, for a flat universe the data vector to consider is $(R, l_A,\Omega_b h^2)$ with
\begin{equation}
 X =\left(
 \begin{array}{c}
 1.74963 \\
 301.80845 \\
 0.02237
\end{array}\right).
\end{equation}
Here $l_A$ is the ``acoustic scale'' defined as
\begin{equation}
l_A = \frac{\pi d_L(z_*)}{(1+z)r_s(z_*)},
\end{equation}
where $d_L(z)$ is the proper luminosity distance and the redshift of decoupling $z_*$ is
given by \cite{husugi},
\begin{equation}
z_* = 1048[1+0.00124(\Omega_b h^2)^{-0.738}]
[1+g_1(\Omega_{m}h^2)^{g_2}],
\end{equation}
\begin{equation}
g_1 = \frac{0.0783(\Omega_b h^2)^{-0.238}}{1+39.5(\Omega_b
h^2)^{0.763}},
 g_2 = \frac{0.560}{1+21.1(\Omega_b h^2)^{1.81}},
\end{equation}
The ``shift parameter'' $R$ defined as \cite{BET97}
\begin{equation}
R = \frac{\sqrt{\Omega_{m}}}{c(1+z_*)} D_L(z).
\end{equation}
$C_{CMB}^{-1}$ in Eq. (\ref{cmbchi}) is the inverse covariance
matrix,
\begin{equation}
C_{CMB}^{-1} = 10^{-8}\left(
\begin{array}{ccc}
1598.9554 & 17112.007 & -36.311179\\
17112.007 & 811208.45 & -494.79813\\
-36.311179 & -494.79813 & 2.1242182
\end{array}\right).
\end{equation}


\end{document}